\documentclass[aps,prl,twocolumn,superscriptaddress]{revtex4}
\usepackage{graphicx}
\usepackage{epstopdf}
\usepackage{amssymb}
\usepackage{amsmath}
\usepackage{color}

\begin{document}

\title{Distributed Imaging for Liquid Scintillation Detectors}

\author{J.~Dalmasson}
\affiliation{Physics Department, Stanford University, Stanford CA 94305-4060, USA}

\author{G.~Gratta}
\affiliation{Physics Department, Stanford University, Stanford CA 94305-4060, USA}

\author{A.~Jamil\footnote{Now at Yale University.}}
\affiliation{Physics Department, Stanford University, Stanford CA 94305-4060, USA}
\affiliation{Erlangen Centre for Astroparticle Physics (ECAP), Friedrich-Alexander-University Erlangen-N\" urnberg, Erlangen 91058, Germany}

\author{S.~Kravitz\footnote{Now at LBNL.}}
\affiliation{Physics Department, Stanford University, Stanford CA 94305-4060, USA}

\author{M.~Malek}
\affiliation{Physics Department, Stanford University, Stanford CA 94305-4060, USA}

\author{K.~Wells}
\affiliation{Physics Department, Stanford University, Stanford CA 94305-4060, USA}

\author{J.~Bentley}
\affiliation{Institute of Optics, University of Rochester, Rochester, NY 14627-0186, USA}

\author{S.~Steven}
\affiliation{Institute of Optics, University of Rochester, Rochester, NY 14627-0186, USA}

\author{J.~Su}
\affiliation{Academy of Opto-Electronics, Chinese Academy of Sciences, Beijing, P.R.China 100094}

\date{\today}

\begin{abstract}

We discuss a novel paradigm in the optical readout of scintillation radiation detectors.   In one common configuration, such detectors are homogeneous and the scintillation light is collected and recorded by external photodetectors.   It is usually assumed that imaging in such a photon-starved and large-emittance regime is not possible.  Here we show that the appropriate optics, matched with highly segmented photodetector coverage and dedicated reconstruction software, can be used to produce images of the radiation-induced events. In particular, such a ``distributed imaging'' system can discriminate between events produced as a single cluster and those resulting from more delocalized energy depositions.  This is crucial in discriminating many common backgrounds at MeV energies.   With the use of simulation, we demonstrate the performance of a detector augmented with a practical, if preliminary, set of optics.  Finally, we remark that this new technique lends itself to be adapted to different detector sizes and briefly discuss the implications for a number of common applications in science and technology.

\end{abstract}

\maketitle

\section{Introduction}

Homogeneous radiation detectors based on the scintillation process have been in use for the last 60~years in a variety of applications, particularly when using liquid organic materials, owing to their simplicity.   Such detectors are relatively cheap to build in very large sizes and, for certain applications, can reach extremely low backgrounds, since the large volume of liquid also shields the inner sensitive volume from external radiation~\cite{KL, detectors}.

It is a commonly assumed that these detectors are fundamentally unsuitable for imaging applications, owing to the ``photon-starved'' regime and uniform angular distribution of the light produced in the scintillation process.   Often this limitation is expressed in terms of Liouville's theorem~\cite{Garwin}, sometimes referred to as theorem of the ``conservation of etendue'' in the field of optics.   This basically states that light emitted over a large angular range ($4\pi$ in the case of scintillation) cannot be imaged through a modest aperture optical system while, at the same time, collecting all (or a large fraction) of the photons produced.  Surely, imaging via appropriate wide angle lenses was central in the operation of bubble chambers~\cite{BubbleChambers}, but in this case external illumination provided sufficient track brightness to produce a conventional image through a lens of finite (and, in fact, relatively small) aperture.  While work on producing scintillators with better light yield is being pursued~\cite{higher_yield_scintillators}, the typical energy yield (defined as the sum of the energy of visible photons over the overall energy deposited) of existing materials can already be as high as $\approx 5\%$, so that improvements in this area are not expected to render conventional imaging possible.

Homogeneous scintillation detectors give up imaging entirely, as the price to pay to collect a large fraction of the emitted photons by maximizing the coverage of photodetectors.   When fast scintillators and photodetectors, typically photomultiplier tubes (PMTs), are used, some crude event localization can be retained by measuring the difference in the times of arrival of the light at each photodetector.  The position resolution of this method is limited by the large value of the speed of light and the finite time resolution of the photodetectors.  More importantly, the emission time of the scintillator, along with the ``late photons'' rescattered in the medium or reflected by detector components, create confusion, drastically limiting the ability to localize events in the volume.  In particular, while for a single energy deposit (with a Gaussian point spread function) the RMS position resolution improves with $\sqrt{N}$, where $N$ is the number of detected photons, the ability to separate two or more spatially distinct sources of light follows a substantially more complex scaling, generally making a classification in terms of multiplicity impossible for homogeneous liquid scintillation detectors.

Yet, each photon arriving at the photodetector surface contains information about its propagation direction although the standard paradigm discussed above cannot capture such information.   It is convenient to analyze the situation in terms of phasespace for the photons that, in this photon-starved regime, can be conveniently treated as individual particles.   Ignoring polarization, each photon is described by six coordinates: its impact position on the photodetector surface $(x,y)$, the time of arrival $t$ (that can be thought of as the third spatial coordinate $n c t$), two directions of propagation, $\theta$ and $\phi$, and the wavelength $\lambda = c n /f = h/p$. Here $f$ is the frequency, $n$ the refractive index of the scintillator, $c$ the speed of light in vacuum, $h$ is Planck's constant and $p$ the momentum of the photon.   Conventional detectors only record $x$, $y$ and $t$.  The additional recording of $\theta$ and $\phi$ allows the three-dimensional reconstruction of the position of the sources ($\lambda$ does not contain information that is relevant here).  This can be achieved by further segmenting the photodetectors and interposing converging lenses in front of them.  Such lenses, if properly designed, convert $\theta$ and $\phi$ into ``local positions'' on the focal array immediately behind them, so that the direction of propagation of the photon becomes available.  Of course, the full information available in this way is distributed over the entire array of photodetectors as, in the photon-starved regime of interest here, too few photons land on any given lens to form an image.   Computer reconstruction is then required to convert the pattern of hits on the photodetectors into three-dimensional images, with a principle similar to that of plenoptic imaging~\cite{Ng}.   The full coverage of lens-photodetector assemblies guarantees an overall light collection efficiency that is similar to that of conventional designs.

 \begin{figure}
\includegraphics[width=3.4in]{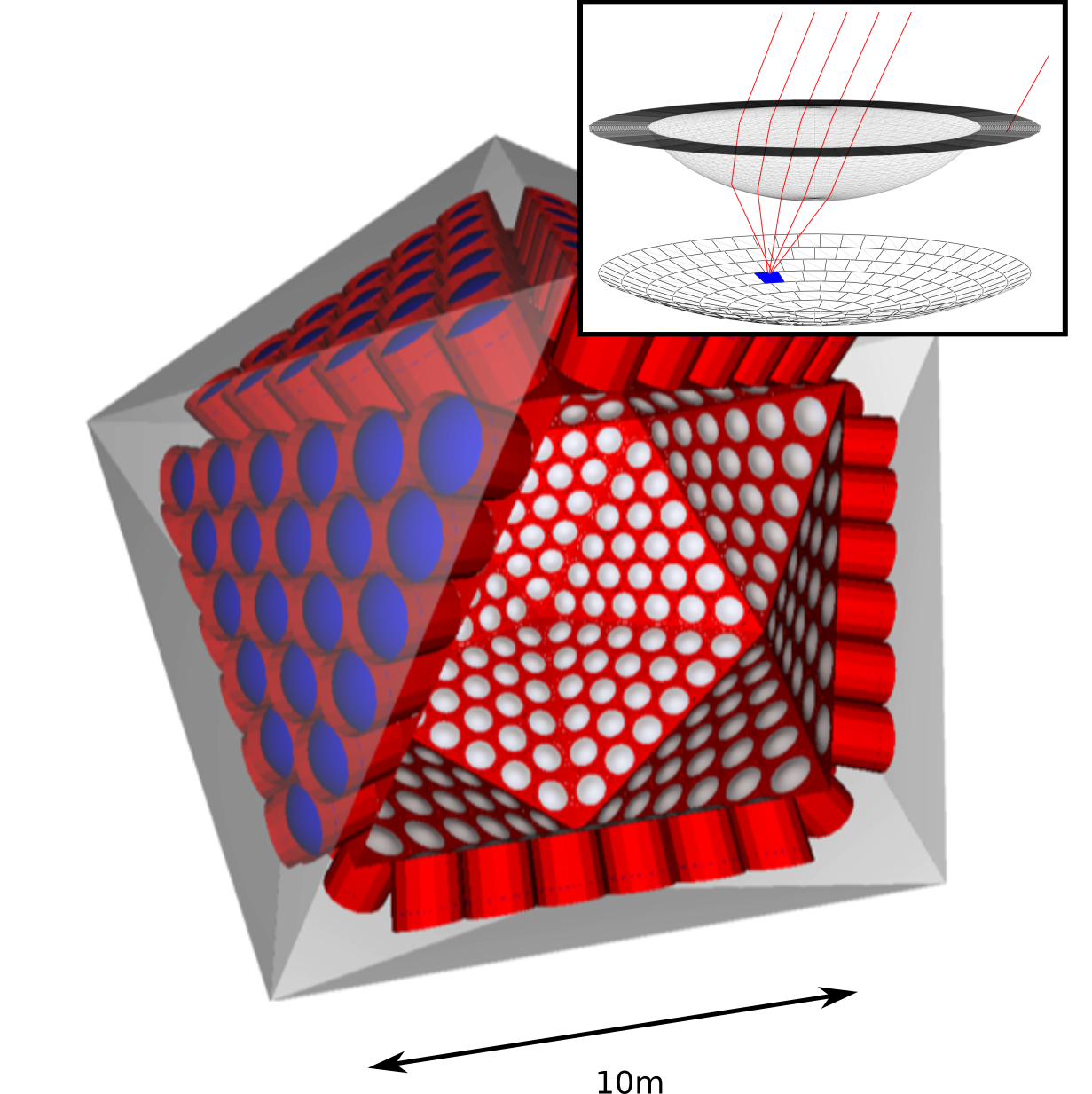}
\caption{Overall geometry of the detector.  For simplicity the baseline concept has lenses and photodetectors arranged on the faces of a regular icosahedron with an inscribed spherical diameter of 14.8~m, similar in size to KamLAND.   In the geometry represented here, one can imagine the large PMTs conventionally used to be replaced by lenses, followed by sub-arrays of smaller photodetectors.  The front of the lenses is shown in white and the back of the photodetectors in blue.  The diameter and number of lenses and, therefore, the multiplicity of the sub-arrays is varied in the optimization discussed in the text.  The inset conceptually shows the details of a lens-photodetector assembly.   The lens directs all rays impinging with a certain angle onto the same readout pixel (within aberrations).  Photons blocked by the pupil limiting the active portion of the lens or the cylindrical baffles between lenses (both shown in red in the main figure) are removed. }
\label{fig:detector_geometry}
\end{figure}

\section{Detector Concept}

While the basic concept described here may appear rather straightforward, some care is needed to demonstrate that realistic optics and event reconstruction can actually provide new information of the kind required to qualitatively improve scintillation detectors.  For the purpose of this early conceptual exploration of the technique, we consider a detector of the size and generic shape comparable to KamLAND~\cite{KL}, simplified by assuming an entirely uniform population of photodetectors (i.e. no provision is made for physical access to the interior of the detector).  In addition, the detector employs only one fluid, i.e. no buffer oil layer is present between the active scintillator and the photodetectors.  Apart from simplifying the detector modeling, this may be practical, since the imaging properties of the system may allow for better fiducialization. The baseline geometry is shown in Figure~\ref{fig:detector_geometry}.  Lenses are arranged in the place of the large PMTs and arrays of smaller photodetectors are placed at the focal surfaces of the lenses. A photon (or a plane wave) impinging on a lens with angles $\theta$, $\phi$ is steered onto a specific element (``pixel'') of the photodetection system, hence measuring the direction of arrival.  In much of the optimization discussed here, the total number of photodetector pixels is held constant at $\simeq 10^5$, while the size and number of the lenses is changed.  With these parameters, the diameter of each photodetector is about 10~cm, consistent with very common PMTs.   

Conceptually, a small number of large lenses, each mapping light onto a larger number of pixels, corresponds to better angular resolution, as this configuration provides higher granularity for the measurement of $\theta$ and $\phi$ (up to the point where lens aberrations start dominating). At the same time, larger lenses correspond to worse spatial resolution, because the system provides no information about the point of impact on the surface of each lens.
   
Conversely, smaller lenses, each mapping light onto an array with fewer pixels, provide worse angular but better spatial resolutions.  In the simplified geometry used here, lenses are arranged on a regular icosahedron with an inscribed sphere of 14.8~m diameter. This is not ideal as rays impinging on lenses at the edges of each face are biased to have larger angles than necessary, but it simplifies the process of parameter optimization, because a change in number of lenses only requires re-tiling of flat (triangular) surfaces.

For simplicity, lenses are circular and photons impinging on the dead regions between them are suppressed (and properly accounted for in estimating the light collection efficiency).  Cylindrical baffles between lenses suppress photons that would land on neighboring focal arrays.  It is conceivable that in a carefully designed detector at least some of these photons would be recovered using lenses with better fill-factor and by accounting for photons landing on contiguous focal arrays, using an algorithm to resolve ambiguities.  Such a more sophisticated design would provide a larger collection efficiency than obtained here.

At an energy of 2~MeV, that is typical in the detection of solar or reactor neutrinos or radioactive nuclear processes, using a conservative scintillator yield of 8000 optical photons/MeV (50\% of anthracene) and the detection efficiency discussed below, the average occupancy on each of the $10^5$ pixels is $<0.04$, if the energy deposit is at the center of the detector.  Therefore, most of the pixels will detect no photons.  Clearly, there is not enough information in any of the individual focal arrays corresponding to lenses to form images.  However, with the angle of incidence information available, each hit can be thought of as a track, analogous to the ionization trail left by charged particles in a time projection chamber~\cite{TPC} or other tracking detector.  Hence the reconstruction of an event will proceed in a manner that is analogous to vertex finding in those detectors.  In addition, the total amount of light recorded will represent the total energy deposited.   Some additional topological information is contained in the time of arrival of the photons at the focal surfaces.  Depending on the timing properties of the scintillator and of the photodetectors, such information, disregarded here for simplicity, may be added to the event reconstruction.  While the position resolution of the detector for point-like energy depositions is studied for different detector parameters, most important is the ability of this technique to image the energy depositions in 3D.  This allows for particle identification, yet maintains most of the simplicity that is typical of scintillation detectors.

Although the focus of this work is on large detectors for rare phenomena, it is expected that the same technique could be applied to smaller size devices, with correspondingly scaled pixel and lens sizes and appropriate photodetector technology.

\section{Individual Lens Design}

The lens design presents unique challenges due to the demanding specifications of both a large numerical aperture and large field of view, required to maximize photon collection efficiency. 
For the same reason, the ratio of the entrance pupil diameter to the diameter of the focal surface should also be as close as possible to unity.
The immersion of the system in the liquid scintillator (with refractive index here assumed to be $n_{\rm scint} = 1.5$) presents a further challenge. However, since the emission spectrum of the scintillator is narrow, typically peaked around 450~nm, the system does not require achromatizing.  Furthermore, the substantial angular size of each pixel means that the lens does not need to be diffraction limited.

\begin{center}
\begin{table}[ht]
{\scriptsize
\begin{tabular}{|l|c|c|}
\hline
 {\bf Parameter}         & {\bf Design 1}  & {\bf Design 2}  \\
\hline
 Number of elements      & 1 (aspherical)  & 2 (spherical)   \\
 $N\! A_{\rm max}$              & 0.62            & 0.64            \\
 $R_{\rm lens}/R_{\rm FA}$ & 0.76         & 1.0             \\
 Angular Resolution      & $4^{\circ}$     & $4^{\circ}$ \\
 $n$                     & 1.98            & 1.98            \\
\hline
\end{tabular} 
}
\caption{Design parameters for the two lens systems used.    Here we provide the maximum values of the numerical aperture ($N\! A_{\rm max}$) and the ratio between the lens radius and the radius of the focal array ($R_{\rm lens}/R_{\rm FA}$).  Note that in Design 1 (Design 2) the focal surface has a radius which is larger than (identical to) the lens radius.   The angular resolution is defined as the RMS of the point spread function of the lens system on the optical axis.  The refractive index of the scintillator is assumed to be $n_{\rm scint} = 1.5$ and optical elements are modeled using high index glass S-NPH2 from the Ohara corp.~\cite{Ohara} with index $n=1.98$ at 450~nm (which is typical of liquid scintillators emission).  In both designs the focal surface is spherical and a pupil of radius $R_{\rm pupil}$ can be used to reduce the numerical aperture of the lens in the process of optimization of the entire detector.  }
\label{tab:lens}
\end{table}
\end{center}

In this work, two different lens systems were designed, with the sole purpose of demonstrating feasibility and providing a practical example of performance to be used in evaluating the power of the technique.  The two systems, with parameters listed in Table~\ref{tab:lens}, are deliberately different: in one case a single, aspherical element is used, while in the other there are two spherical elements.  The numerical aperture is defined as $N\! A = n_{\rm scint} \sin{[\arctan({R_{\rm pupil}/f_{\rm eff}})]}$, where $R_{\rm pupil}$ is the radius of the entrance pupil and $f_{\rm eff}$ is the effective focal distance.   While in Design 2 the diameter of the lens is identical to that of the focal surface, Design 1 requires a focal surface that is larger than the lens, resulting in a less optimal packing of the lenses and therefore lower light collection efficiency. In the optimization process for the entire detector described below, $R_{\rm pupil}$ is varied, resulting in $N\! A\le N\! A_{\rm max}$.   

A spherical focal surface is used in both cases, as this may be feasible for photodetectors with independent pixels (e.g. arrays of photomultiplier tubes) and it simplifies the initial optics design.  The curvature of the focal array is optimized to minimize the number of optical elements and hence increase the efficiency of light collection over the full field of view. To further help maximize the overall efficiency, the angles of incidence and refraction on the lenses were kept to a minimum, reducing reflections at the boundaries.    Lenses with refractive index $n>n_{\rm scint}$ (high index) and $n < n_{\rm scint}$ (low index) materials were tested. The use of low index elements (e.g. air elements) did not prove to be beneficial in this preliminary study.   Given the modest angular resolution requirements, moldable glass and plastics appear to be plausible materials, although the requirement on the high refractive index may constrain the choices.  Specific manufacturability and cost considerations are disregarded here.   Lastly, hermetically sealed designs, alternating low index regions with high index elements were compared to immersed designs but were eventually discarded due to their added complexity. 

\begin{figure}
\includegraphics[width=3.4in]{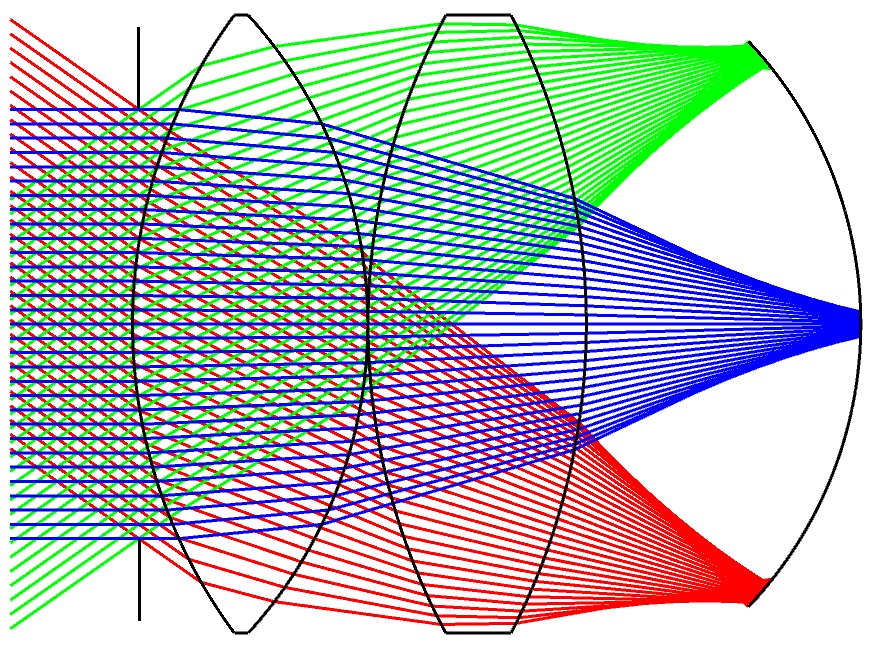}
\caption{Diagram of a Design~2 optical assembly with on-axis and full field rays.  Also shown is the pupil.   The parameters of the design are given in Table~\ref{tab:lens}.  No linear dimensions are provided, since these are changed in the course of the optimization while maintaining all angles constant.}
\label{fig:Lens}
\end{figure}

Design 2, with more optical surfaces, results in smaller angles of incidence and refraction, hence achieving smaller light losses than the single-lens Design 1.    For instance in the case of 538~pixels/lens and 200~lenses with maximum numerical aperture Design~1 (2) achieve 8\% (20\%) photon collection efficiency.   For this reason, only results from Design~2, shown in Figure~\ref{fig:Lens} with on-axis and field rays, are described in the rest of this paper.

The detector optimization proceeds by varying the size of the lenses and the number of pixels behind each lens.  This is achieved with a similarity transformation applied to each lens assembly, keeping the size of each pixel, and hence the total number of pixels in the entire detector, approximately constant.  Here and in the following ``lens assembly'' refers to the combination of lens(es), pupil and focal array, shown in Figure~\ref{fig:Lens}.

\section{Event Simulation and Reconstruction}

Detector performance is studied using the {\ttfamily CHROMA} ray-tracing package~\cite{CHROMA} that runs on Graphics Processing Units (GPUs) and is substantially more computationally efficient than CPU-based software.   {\ttfamily CHROMA} also includes an interface to {\ttfamily GEANT4}~\cite{GEANT4}.  An ``event'' is the overall energy deposit in the scintillator from some interaction, resulting in a 3D pattern of scintillation light that may originate from one or more ``sites'' (or even from a continuous distribution of sites, as in the case of an ionized track).   Scintillation emission is considered monochromatic with the exact value of $\lambda$ being irrelevant from the point of view of the simulation.   Photons are always emitted isotropically at each site, starting from the $8000$~photons/MeV yield and immediately applying a detection quantum efficiency of 33\%, which is realistic for modern photocathodes and conservative for many solid state photodetectors.  Self-absorption in the scintillator material is not considered.  To estimate the confusion arising from dark current hits, we assume a photocathode dark rate of 10~cm$^{-2}$s$^{-1}$ (reasonable at 10~$^\circ$C) and an integration time window of 50~ns.  This would result in a total rate of $\sim 3$~event$^{-1}$, too small to be of concern and hence not considered further.  Ray-optics effects are properly accounted for by the simulation, including some confusion derived from reflections and the loss of some photons.  

As mentioned, most of the study is carried out by setting the total number of pixels to $10^5$.  Some special runs were performed with a larger or smaller number of pixels.   Of course, as the number of pixels increases so does the cost. 

In the detector optimization, the number of lenses per face of the icosahedron (and hence the total number of lenses in the detector) is varied, resulting in a different number of pixels per lens assembly; the product of the two being constant (e.g. at $\sim 10^5$).  In order to assess the position-dependence of the detector performance, for most of the runs, events, whether point-like, dummy photon sources or physics generated by {\ttfamily GEANT4}, are produced at random positions in two different regions of scintillator. These regions are a sphere of 1~m radius in the center of the detector and a spherical shell with radii between 3 and 4~m.

\begin{figure}
\includegraphics[width=3.4in]{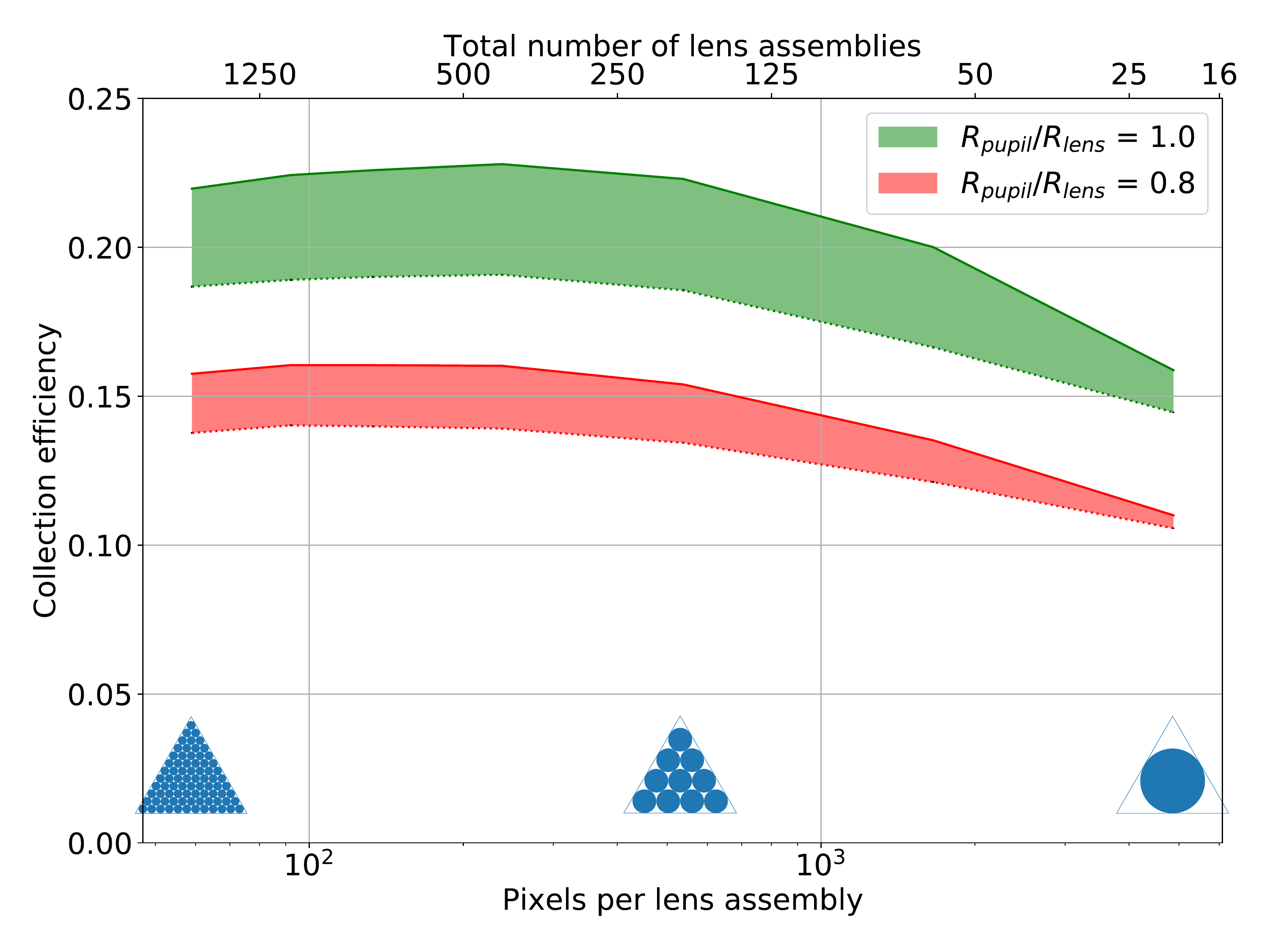}
\caption{Photon collection efficiency comparison for different configurations of detector based on Design 2. 
The two colored bands represent $R_{\rm pupil}/R_{\rm lens}$ of 1.0 (green) and 0.8 (red). The upper and lower bounds of each band refer to events produced randomly in a sphere of radius 1~m in the middle of the detector and in a spherical shell of radius between 3 and 4~m, respectively. The different detector configurations are expressed in terms of number of pixels per lens assembly (bottom scale) or number of lens assemblies in the detector (top scale). The configuration of lenses on one of the 20 faces of the icosahedron is visually illustrated in three cases, for reference. 
}
\label{fig:eff_comp}
\end{figure}

The photon collection efficiency (including quantum efficiency) is shown in Figure~\ref{fig:eff_comp} for Design~2 and two different pupil sizes.  The detector configuration, as just described, is identified in terms of pixels per lens assembly (bottom scale) or total number of lens assemblies in the entire detector (top scale).   The width of the two bands spans the efficiency of the detector for events produced in the central 1~m radius and in the spherical shell mentioned above.   The broad maxima indicate that, from the point of view of light collection efficiency, there is substantial flexibility in the detector design optimization.  The efficiency is also relatively independent from the location of the event, although radius-dependent corrections to the energy will be required, as is typical in these large detectors. 

The performance evaluation proceeds in two stages, as it would in a real detector: a calibration stage, where a model of the detector response to optical photons is generated, and a reconstruction stage, where an algorithm utilizing such a detector response is applied to individual events to determine their properties. In the model employed for the detector response, a ``hit'' (detected photon) on a given pixel $i$ maps onto a most probable three-dimensional direction of origin ${\hat{n}_i}$ for that photon, along with an associated angular uncertainty $\delta\alpha_i$.  This can be visualized as a cone-shaped ``road'' truncated by the aperture of the lens and expanding with distance $z_i$ along direction ${\hat{n}_i}$, with a half-aperture given by $R_{\rm lens} + \sigma_i$, where $\sigma_i = z_i \delta\alpha_i$. The calibration process then consists of generating the values ${\hat{n}_i}$ and $\delta\alpha_i$ for each pixel $i$ in the detector. These parameters are determined by generating photons randomly, with a uniform distribution, and isotropically throughout the detector, then considering only those photons which hit pixel $i$. The mean direction to the origin for these photons is then calculated and assigned to ${\hat{n}_i}$, and the standard deviation of the projections perpendicular to ${\hat{n}_i}$, $\sigma_i$, calculated by adding in quadrature the values on the two axes of projection. This process requires the knowledge of the position of origin of the photons, which in a real detector could be accomplished using a movable calibration source, as is customary~\cite{calibration}.  

After this calibration is completed, there are two major reconstruction goals, corresponding to different algorithms: 1) find the position of one or more distinct sites which are sufficiently separated (resolved) and 2) decide, in a statistical way, whether the light in an event is produced at a single or at multiple, unresolved sites.  While it should be possible to further image events in terms of more complex topological configurations, here we concentrate on differentiating between ``single-site'' and ``multi-site'' events, as this simple classification is of primary concern in the discrimination between electrons, gamma-rays and neutrons at MeV energies.   It is likely that algorithms both in 1) and 2) could be replaced by machine-learning techniques, plausibly resulting in better performance. However, for this preliminary study we prefer to employ more classical methods, so that the results are conservative and easily understood from the geometry of the problem.   

The first reconstruction goal is solved using a standard tracking algorithm~\cite{LHC_tracking} that proceeds iteratively by first trying to find the position of a site through a weighted least-squares optimization, with weights corresponding to the probability of a hit originating from that site (rather than another). Once a site position is determined, hits associated to it are removed, and the algorithm is repeated with the remaining unassociated hits, until all hits are associated to a site. The optimization function used is $O(\mathbf{v}) = \sum\limits_{i=1}^n w_i(\mathbf{v}) \chi^2_i(\mathbf{v})$, where the sum is performed over all roads $i$, $\chi^2_i(\mathbf{v}) = d^2_i(\mathbf{v})/\sigma^2_i$ is the distance from road $i$ to the vertex at position $\mathbf{v}$, normalized by $\sigma_i$ (the width of road $i$ at the vertex position), and $w_i(\mathbf{v})$ is the weight for a given hit. The weight function $w_i(\mathbf{v}) = 1/\left(1+\exp\left((\chi_i^2(\mathbf{v})-\chi_c^2)/2T\right)\right)$ assigns higher value to roads with smaller $\chi^2_i$ values, but in a way that is more forgiving for higher values of the ``temperature'' parameter $T$, used in a simulated annealing algorithm. Reducing $T$ as the vertex position gets closer to its optimum allows for inclusion of many roads during the initial stages, when the vertex position is fairly uncertain, while rejecting those with $\chi_i>\chi_c$ at convergence, when $T$ is small.

\begin{figure}
\includegraphics[width=3.4in]{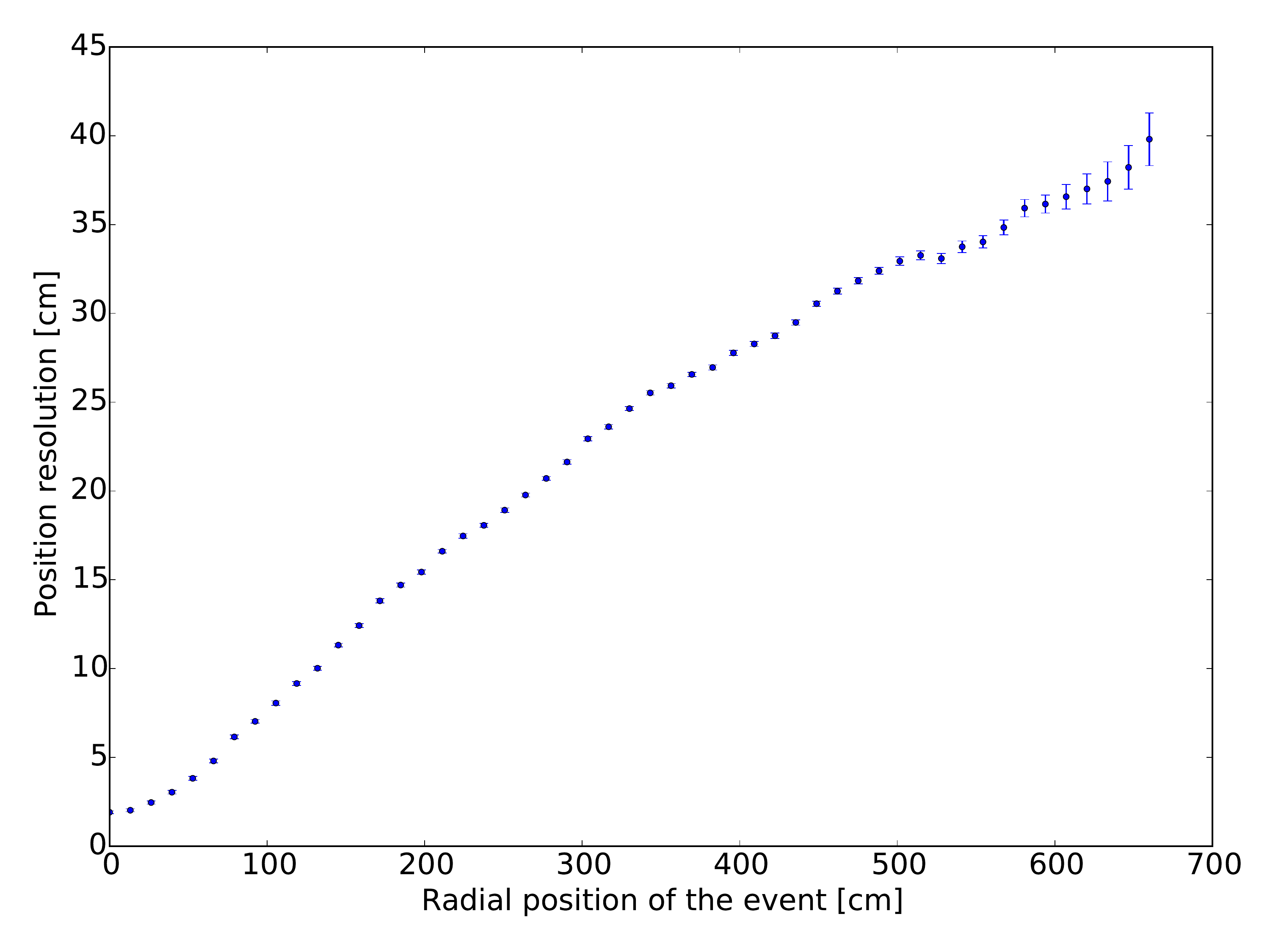}
\caption{Position resolution for single-site energy deposition of 2~MeV at different radii, using a configuration with 538~pixel per lens, 200 lenses and $R_{pupil}/R_{lens}=0.8$. The resolution is defined as the mean distance between the simulated position and the position reconstructed by the algorithm. Each point is obtained by averaging 100 events seeded at the same radial distance but otherwise uniformly distributed. The error bars represent the standard error on each point and include statistical uncertainties only.}
\label{fig:singlesite}
\end{figure}

Figure~\ref{fig:singlesite} shows the position resolution for a 2~MeV single-site deposit at various radii in the detector, obtained for the baseline configuration with 538~pixels per lens and 200~lenses. We note that such resolution degrades appreciably above 5~m radius, but this is a substantially larger radius than is usually considered for fiducialization.  In addition, it is likely that the situation at large radii is exacerbated by the icosahedral shape of the detector surface.  

In order to test the ability of the detector to classify events as single- versus multi-site, we directly compare simulated energy depositions from electrons and $\gamma$-rays with the same energy.   While these two types of events are not strictly single- or multi-site, they represent the proper figure of merit required in most low energy applications, and they are basically indistinguishable by a conventional liquid scintillation detector.  Both types of events are produced in {\ttfamily GEANT4} with 2~MeV total energy in the two concentric regions described above.  Events of lower (1~MeV) and higher (10~MeV) energy are also simulated for additional tests.

After generating an e$^-$ or $\gamma$ event, the calibration data is used to reconstruct a road for every hit. The algorithm then calculates the most probable distance of closest approach for each pair of roads.  These distances, $d_{ij}$, have uncertainties $\sigma_{ij} = \sqrt{\sigma_{i}^2 + \sigma_{j}^2}$ where $\sigma_{i}$ and $\sigma_{j}$ are the transverse widths of the two roads $i$ and $j$ at the position of closest approach.  For each event, the weighted distances $d_{ij} / \sigma_{ij}$ are then histogrammed.   A ``e$^-$ template'' is produced by the histogram filled with the average of 500 electron-type events.  In a subsequent run, the $\chi^2$ is computed between the histogram of e$^-$ or $\gamma$ events and the ``e$^-$ template''.  The cumulative distributions of $\chi^2$ for all e$^-$ and $\gamma$ events are then formed and a hypothesis test estimates the likelihood of electrons and $\gamma$'s as being properly classified.  As a figure of merit we report the probability of rejecting $\gamma$'s, when the efficiency for properly classifying electrons is 80\%.  In a typical run 500 e$^-$ or $\gamma$ are generated (in addition to the events used to build the template) and the uncertainties are derived as asymptotically normal from the e$^-$ cumulative distribution and propagated on the $\gamma$ cumulative distribution using the delta method~\cite{stat}. This procedure is cross-checked with a bootstrap technique on the Monte Carlo event samples~\cite{Efron}.

\begin{figure}
\includegraphics[width=3.4in]{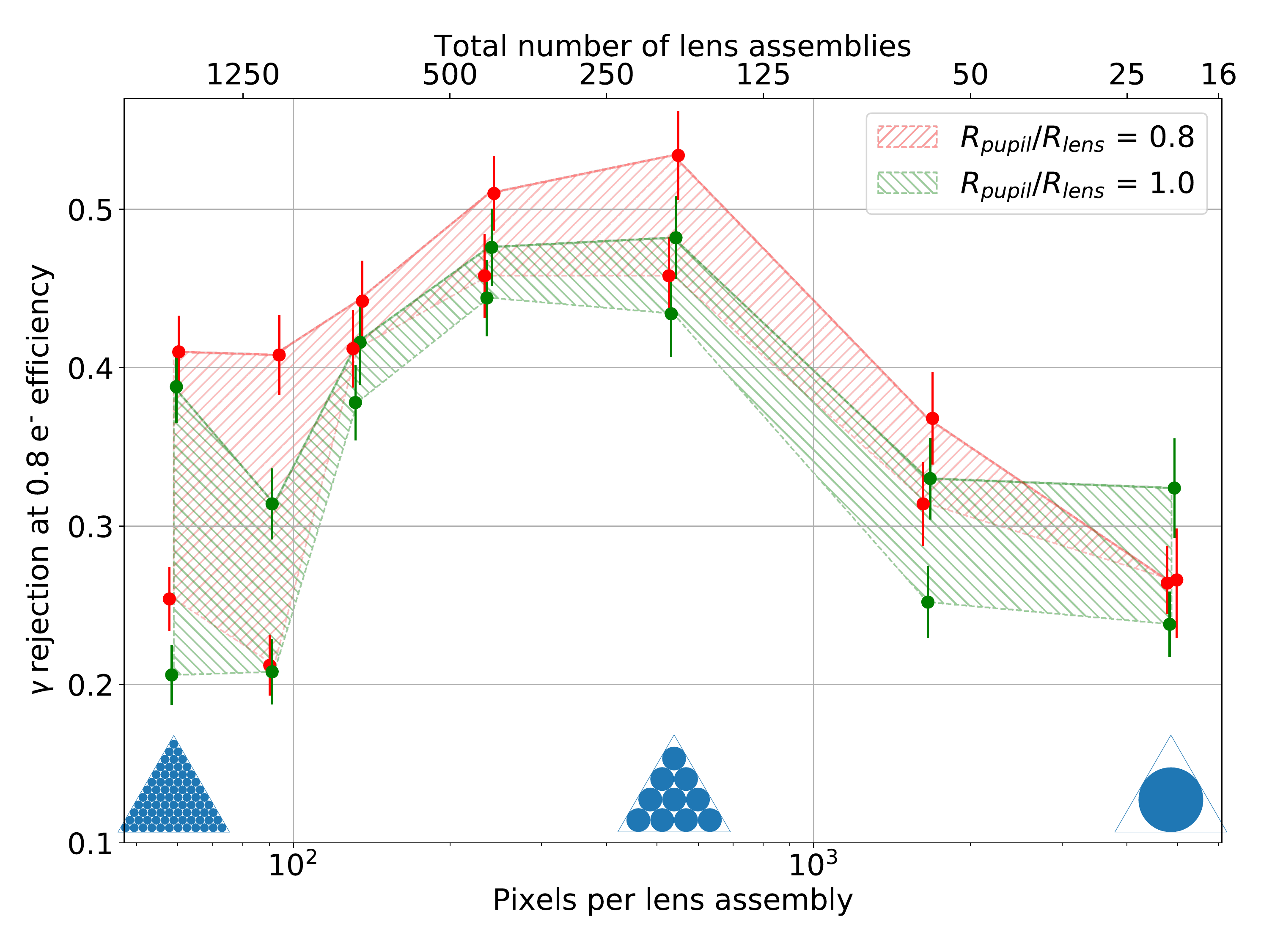}
\caption{Detector performance for a variety of configurations and two different pupil sizes.  The performance is stated in terms of the $\gamma$ rejection factor for 80\% electron efficiency at an energy of 2~MeV, as described in the text.   The data points are slightly shifted horizontally to avoid overlaps and improve visibility.   The width of the two bands is bounded by the performance in the center region of the detector and in the wider spherical shell, as explained in the text.   The different detector configurations are expressed in terms of number of pixels per lens assembly (bottom scale) or number of lens assemblies in the detector (top scale).  The configuration of lenses on one of the 20 faces of the icosahedron is visually illustrated in three cases, for reference.  Error bars represent the variance of each point.}
\label{fig:main_result}
\end{figure}

\section{Results}

The ability of the detector to reject $\gamma$-rays while retaining 80\% of the e$^-$ interactions is shown in Figure~\ref{fig:main_result} as a function of the detector configuration.  The discriminating power derives from the more dispersed nature of $\gamma$ events, where the energy is distributed in space by Compton scattering.  As expected, the best performance balances the fine angular resolution from many pixels per lens against the spatial resolution improvement for smaller lenses.

The analysis predicts this trade-off to produce the best performance with a few hundred pixels per lens assembly and a total of a few hundred lens assemblies.  This result appears to have little dependence on the location of the energy deposition in the detector volume.   As already mentioned, we expect that the several simplifications made will limit the performance of the detector.  We also note that the ability to identify the $\gamma$'s is not only important to reject backgrounds, but can also be used to directly measure the background in the detector.  This feature is entirely new for scintillation detectors.

\begin{figure}
\includegraphics[width=3.4in]{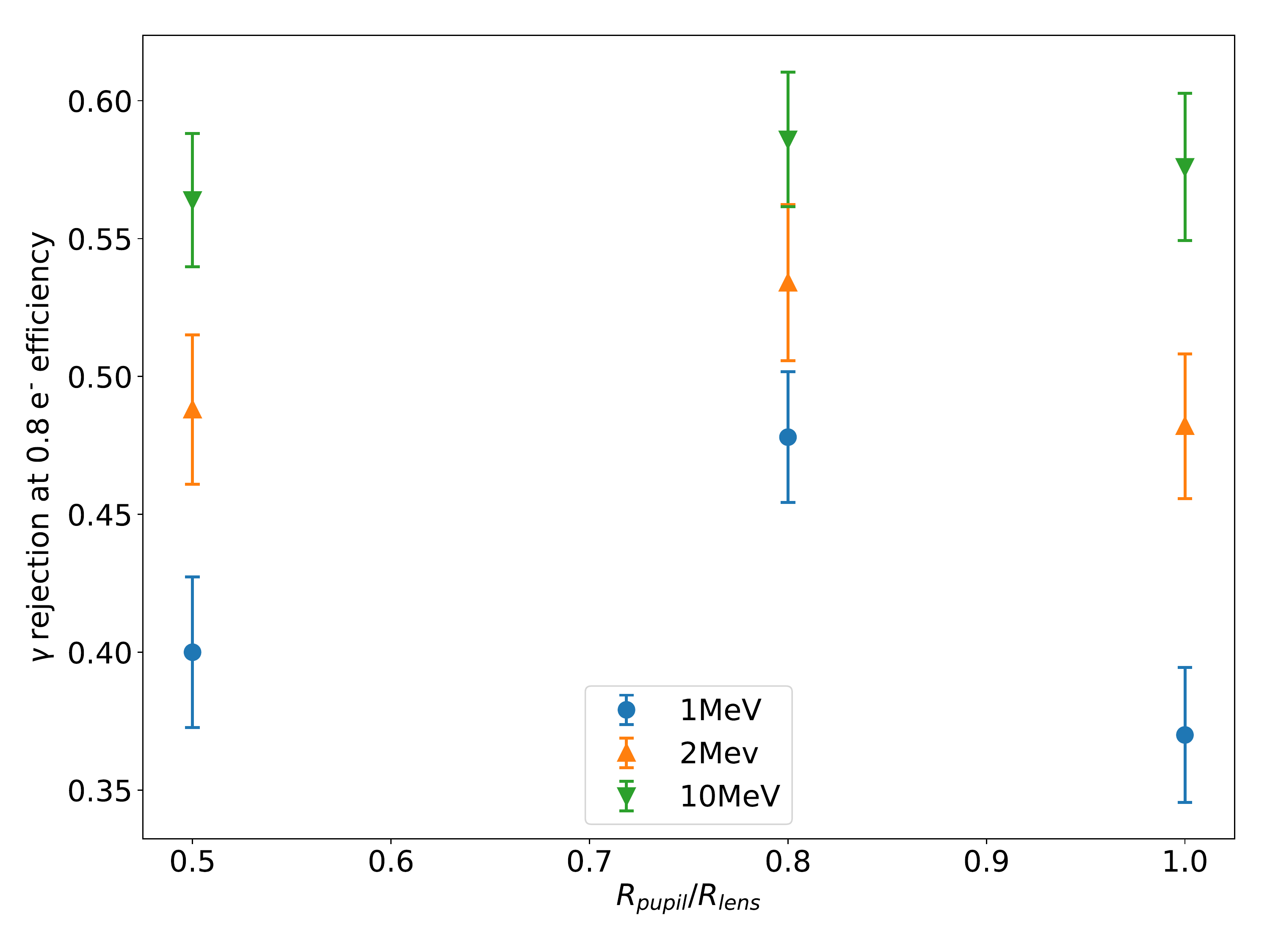}
\caption{Overall detector performance, for the detector configuration with 538 pixel per lens assembly and 200 lens assemblies, as a function of $R_{\rm pupil}/R_{\rm lens}$. Only events generated in the central 1~m radius sphere are used. The performance represents the $\gamma$ rejection factor for 80\% electron efficiency at an energy of 2~MeV, as described in the text. Here, results for particles of 1~MeV and 10~MeV of energy are also shown.}
\label{fig:MST}
\end{figure}

Figure~\ref{fig:MST} shows the $\gamma$ rejection efficiency for one of the better configurations (538 pixel per lens assembly and 200 lenses assemblies) and three different pupil sizes.  The trade-off between lens aberrations and photon statistics is evident, particularly for the 2~MeV case, that is used as the primary energy in this work. A modest pupil ratio of 0.8 has better performance than the full lens (1.0) because of the reduction in aberrations.  However, further reduction of the pupil results in a degradation of the performance because of the lower photon statistics.  The latter effect is substantially less pronounced at 10~MeV, where more photons are available.  A better lens design is expected to improve performance by allowing for better quality imaging with larger pupil size.  This is also expected to improve the energy resolution of the detector, which is not analyzed here but scales, as usual, with $1 / \sqrt{N}$, where $N$ is the number of photons collected.  At the same time the total number of pixels and, ultimately, the spatial structure of e$^-$ and $\gamma$ (and neutron, if interesting) events set a natural bound on how much information can be extracted from the system.

\begin{figure}
\includegraphics[width=3.4in]{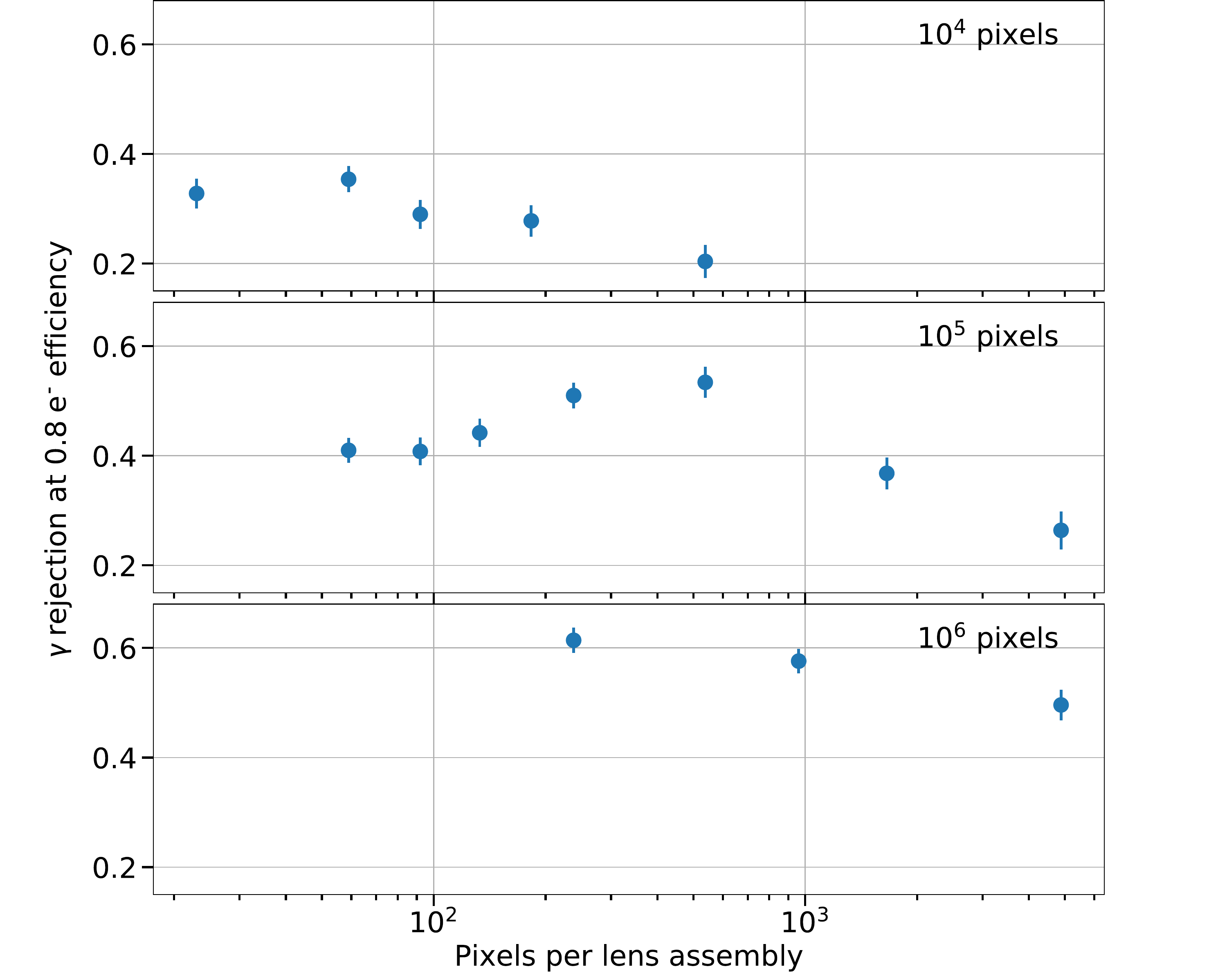}
\caption{$\gamma$ rejection efficiency for 80\% e$^-$ retention and different detector configurations, now including $10^{4}$, $10^{5}$ and $10^{6}$ pixels in the entire detector.  Each panel represents the case of a different number of pixels, while the abscissa indicates the detector configuration, in terms of number of pixels per lens assembly.  Here all configurations have $R_{\rm pupil}/R_{\rm lens} = 0.8$ and the events are all seeded in the central 1~m radius part of the detector. 
}
\label{fig:DE_vs_Dd}
\end{figure}

To further investigate the role of the overall pixel granularity, we have performed runs in which the total number of pixels in the detector was altered from the default of $10^5$ down to $10^4$ and up to $10^6$. The result of this study is shown in Figure~\ref{fig:DE_vs_Dd}, which gives the $\gamma$ rejection efficiency for each configuration and each total number of pixels.  The rejection efficiency is substantially degraded for $10^4$ pixels, while at $10^6$ pixels any configuration achieves a similar, saturated rejection efficiency.  We surmise this to be due to aberrations in the lenses that do not allow the detector to take advantage of the larger number of pixels.  This can be confirmed by a simple back of the envelope comparison between the angular extent of a pixel and the angular resolution of the lens from Table~\ref{tab:lens}.

\section{Conclusions}

We have described a novel technique to produce images in photon-starved environments, with light sources of very large emittance. This is the typical case for scintillation detectors of ionizing radiation, where traditionally it has been assumed that imaging is impossible.  We wish to call this new technique, which is analogous to plenoptic imaging, ``distributed imaging,'' as the information required to form the image can only be obtained by associating the data collected by many lens systems arranged in a way to entirely surround the source of light.     

In order to demonstrate the power of the technique in a realistic situation, we have simulated a large liquid scintillator detector, similar in dimensions to KamLAND, and shown that a substantial discrimination between single electrons and $\gamma$ rays can be obtained with a rather simple reconstruction algorithm. We have also performed a first optimization of the overall performance tradeoff between position and angular resolutions.  For an energy of 2~MeV, typical for a number of experiments involving nuclear reactions, half of the events produced by $\gamma$-rays are properly tagged by this technique, maintaining 80\% efficiency for recognizing electron events.  This is comparable to the performance obtained with far more complex detectors, which directly read out the ionization.  A number of simplifications made in the geometry, as well as the preliminary nature of the design of the lens system, suggest that further performance improvements are possible.  Better performance is also expected from reconstruction algorithms based on machine learning.  Such techniques were deliberately not used in this first analysis in order to gain a better understanding of the physical principles at play.   Timing was not included in the work presented here, and its addition may provide some performance improvement, especially for the position resolution of distinct sites.

While the required segmentation of the photodetector system increases the cost with respect to conventional, non-imaging, scintillation detectors, the topological information is of great value in separating signals from backgrounds at low energy or in providing full tracking information for $\sim$GeV interactions.  We note that a system with $10\times 10$ arrays of 1.5-inch diameter PMTs in the place of single 20-inch diameter ones is likely to increase the photodetector cost by only a factor of $\sim 5$, since 1.5-inch PMTs (e.g. Hamamatsu~\cite{HPK} R11102) are priced at about 5\% of the cost of 20-inch ones (e.g. Hamamatsu R3600-2 or R7250).   In addition, rapid progress in the area of photodetector development may further mitigate the cost issues, particularly as time resolution is not essential in this method.  

Further optimization of the lens system with specific attention to the overall light efficiency is expected to lead to improved results. To allow flexibility in photodetector choice (e.g. large area PMTs with positions-sensitive anodes) future designs with a planar focal surface also need to be investigated. In addition, a more robust manufacturability analysis would benefit from further investigation into ``hybrid'' designs, involving hermetically sealed components, air-core and low-index plastic lenses.

Finally, we note that, unlike in the case of position reconstruction from timing, where the resolution is a constant related to the combination of the time structure of the scintillator emission and the speed of the photodetectors (and the large value of $c$), the resolution achievable here only depends on the size of the pixels and the quality of the lenses employed.  Thus the principles described have applications beyond those of large detectors for rare events. In particular, the present technique scales favorably when the detector dimensions decrease, since maintaining constant the total number of pixels (while reducing their size) results in improved spatial resolution.   Segmented-anode PMTs or solid state photodetectors may already be available for the use in smaller systems, although the development of new, highly segmented photodetectors not requiring particularly good timing properties is also a possibility.  Hence we expect that this method could be optimized for applications in smaller $\gamma$-ray and neutron detectors for various areas of science and technology.  For instance, different algorithms may allow for the directional detection of $\gamma$s or fast neutrons, by tagging, respectively, energy deposits in Compton scattering and the first steps in the thermalization process. 

Distributed imaging may substantially extend the power of scintillation detectors that are intrinsically simple and relatively inexpensive.  

\begin{acknowledgments}
We wish to thank 
N.~Bareket (Abbott Medical Optics), P.~Hanrahan, B.~Macintosh, D.~Palanker and C.~Pellegrini (Stanford University) for stimulating discussions on optics, J.~Klein (University of Pennsylvania) for an introduction to Chroma, J.~Detwiler (University Washington) for a refresher on event localization in KamLAND and G.~Walther (Stanford University) for a discussion on statistics.  GG would like to particularly thank J.~Learned (University of Hawai'i) for pointing out, in some remote past, the connections between Liouville's theorem and scintillation optics.  The original idea presented here was triggered, in part, by trying to better explain imaging optics in a freshman physics class.   This work was supported by seed funds from Stanford University.
\end{acknowledgments}

\end{document}